# A statistical test to reject the structural interpretation of a latent factor model

Tyler J. VanderWeele
*Harvard University, Cambridge, MA, U.S.A.*

Stijn Vansteelandt
*Ghent University, Ghent, Belgium*

**Summary**. Factor analysis is often used to assess whether a single univariate latent variable is sufficient to explain most of the covariance among a set of indicators for some underlying construct. When evidence suggests that a single factor is adequate, research often proceeds by using a univariate summary of the indicators in subsequent research. Implicit in such practices is the assumption that it is the underlying latent, rather than the indicators, that is causally efficacious. The assumption that the indicators do not have effects on anything subsequent, and that they are themselves only affected by antecedents through the underlying latent is a strong assumption, effectively imposing a *structural* interpretation on the latent factor model. In this paper, we show that this structural assumption has empirically testable implications, even though the latent variable itself is unobserved. We develop a statistical test to potentially reject the structural interpretation of a latent factor model. We apply this test to data concerning associations between the Satisfaction-with-Life-Scale and subsequent all-cause mortality, which provides strong evidence against a structural interpretation for a univariate latent underlying the scale. Discussion is given to the implications of this result for the development, evaluation, and use of measures and for the use of factor analysis itself.

## 1. Introduction

The model underlying classical test theory and much of measure construction often assumes an underlying univariate latent variable which itself gives rise to, or causes, various observed indicators (DeVellis, 2016; Price, 2016). These indicators themselves form the empirical bases of the measures that are constructed. Various psychometric tests are available to evaluate the adequacy of this measurement model and to assess whether a single univariate latent adequately captures the covariance structure among the set of observed item responses or indicators (Thompson, 2004; Comrey and Lee, 2013; Kline, 2014; Brown, 2015). When the evidence seems to indicate that a unidimensional factor is sufficient, the indicators are then typically combined, often simply as a mean of their values, to form a measure that is then used in subsequent research. The measure is thought to be an adequate assessment of the underlying latent variable that corresponds to the relevant construct, that is of theoretical interest and worthy of empirical investigation. The measure will then typically be used in subsequent research to study the causes that might give rise to the phenomenon relevant to the construct under consideration, and also causal relations with other outcomes.

However, when used in this way, a subtle implicit supposition is made which is often overlooked. From the measurement model fitting reasonably well, it is subsequently assumed that it is in fact the supposed underlying latent variable that is causally efficacious (Bollen, 1989; Sánchez, 2005). The individual indicators are assumed to be effectively causally inert, and it is thus the measure, imperfectly but appropriately representing the latent, that is sufficient for use in causal research. In such reasoning, however, an unwarranted leap in logic is in fact made. From the covariance of the individual indicators fitting a univariate latent covariance structure well, it does not follow that it is the supposed underlying univariate latent that is causally efficacious and that the indicators are not. The univariate latent measurement model fitting well is in fact entirely consistent with each indicator having separate distinct causal effects on subsequent outcomes, and with these effects being considerably different from each other across indicators. In this paper, we show that these questions are in fact subject to empirical investigation. It is demonstrated that the structural interpretation of the latent factor model – that it is the supposed univariate latent rather than its indicators that are causally efficacious – imposes assumptions that are sufficiently strong so as to give rise to empirical implications that can be tested, and rejected. We make use of these empirical implications to develop statistical tests that can lead to the rejection of the structural interpretation of a latent factor model. We illustrate this test with empirical data concerning associations between the Satisfaction with Life Scale (Diener et al., 1985) and subsequent all-cause mortality to examine whether an underlying univariate "life satisfaction" latent variable with a structural interpretation is reasonable.

    It should of course be noted that not all measure construction or latent variable modeling proceeds with an assumption of univariate latent variable. A large methodological literature exists concerning model selection, and on identification procedures for using observable model implications of latent variable models for model selection and causal discovery (Glymour and Spirtes, 1988; Bollen, 1989; Spirtes et al., 2000; Silva et al., 2006; Sullivant et al., 2010; Kummerfeld et al., 2014; Gignac, 2016; Kummerfeld and Ramsey, 2016; Mansolf and Reise, 2017). Our intent here is not a general critique of the latent variable and structural equation modeling literature. Rather, our critique is focused on the relatively common practice of simply examining a basic factor model for evidence of unidimensionality, and then thereafter presuming the latent factor model is structural. The empirical implications and the tests we develop in the sections that follow focus on this case. However, if a structural interpretation of a univariate latent factor model is rejected, then additional model selection and identification literature of course becomes yet further relevant. With regard to the present paper, after laying out the empirical constraints implied by a univariate structural latent factor model and the corresponding statistical tests, we then further discuss the implications of this result, and of the often inappropriate assumption of a structural interpretation, for the development, evaluation, and use of measures and for the use factor analysis itself.

## 2. Latent Factors Models and Empirical Implications

    The classical model used in much measurement theory and scale development, sometimes also referred to as a "reflective model", presupposes an underlying continuous latent variable $\eta$ that gives rise to continuous indicators or measurements $(X_1, \ldots, X_d)$ as in Figure 1.

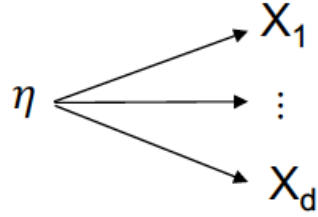

Fig. 1. Basic latent factor model with latent $\eta$ and indicators $(X_1, \ldots, X_d)$.

In practice, the indicators $(X_1, \ldots, X_d)$ are often not strictly continuous but may arise from subject responses from a likert scale and are assumed approximately continuous. After standardization so that each indicator $X_i$ has mean 0 and variance 1, which is often done for comparability and ease of interpretation, it is then often assumed that each indicator $X_i$ is given by a linear function of the latent variable $\eta$ plus a mean-zero random error $\varepsilon_i$ independent of $\eta$:

$$X_i = \lambda_i \eta + \varepsilon_i \tag{1}$$

where $\lambda_i$ are generally assumed unknown and where random errors $\varepsilon_i$ are often, but not always, assumed normally distributed and independent of one another. Throughout, as is common, we will assume that the coefficients $\lambda_i$ are non-zero for all of the indicators, $(X_1, \ldots, X_d)$, since, otherwise, any indicator for which $\lambda_i=0$ would be irrelevant for the latent variable $\eta$ supposedly corresponding to the construct of interest and would thus be omitted. The model above forms the basis of much psychometric measure evaluation (DeVellis, 2016; Price, 2016). However, after this evaluation is complete, the measures that are used in practice are generally just some univariate function of the indicators, $f(X_1, \ldots, X_d)$. Let $A = f(X_1, \ldots, X_d)$ denote the measure that is eventually employed. When the indicators are on the same scale, often the mean of the indicators is used. The function of the indicators is meant to be an imprecise measure, subject to error, of the underlying latent variable $\eta$ that corresponds to the psycho-social construct of interest. It will then often be of interest to assess the relationship of this measure with various other important outcomes. This is what is typically done in structural equation models (Bollen, 1989; Sánchez, 2005).

However, the model in equation (1) is entirely consistent with different sets of causal relationships with some outcome of interest $Y$. Contrast the relationships in Figure 2a and Figure 2b. On the one hand it is possible that it is the supposed underlying variable $\eta$ that has a causal effect on the outcome $Y$ and that the indicators are causally inert as in Figure 2a. On the other hand, it is possible that it is the individual indicators $(X_1, \ldots, X_d)$ that each exert causal effects on the outcome $Y$ as in Figure 2b. Importantly, both of these causal structures are entirely consistent with the measurement model in equation (1) and Figure 1, as are other structures as well such as if both the latent variable $\eta$ and the indicators $(X_1, \ldots, X_d)$ affected $Y$.

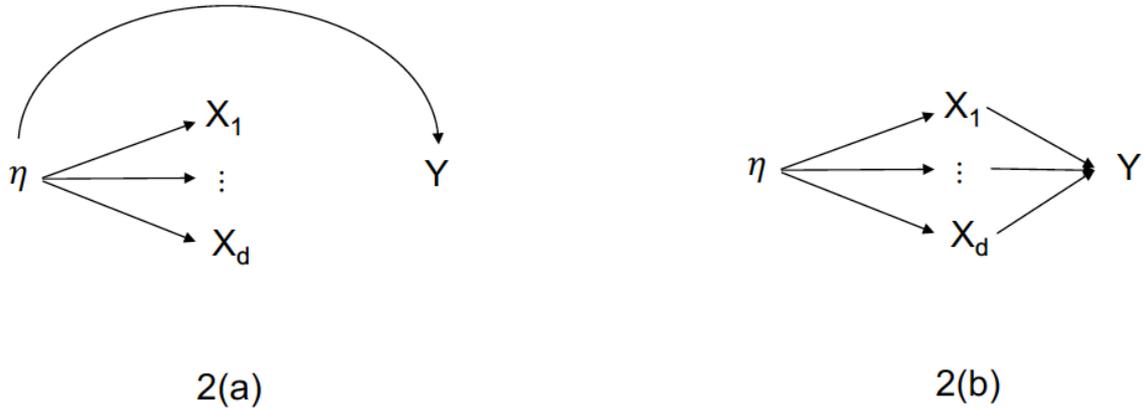

Fig. 2. (a) Structural latent factor model with latent $\eta$ causally efficacious for outcome $Y$; (b) basic latent factor model with indicators $(X_1, \ldots, X_d)$ causally efficacious for outcome $Y$.

Similarly, consider a randomized trial of some treatment $T$ in which we are interested in whether the treatment $T$ might have causal effects relevant to the construct or phenomenon under study. In practice, we would often compare the average value of our measure $A$, across arms of the randomized treatment $T$ and assess causal effects as $E(A|T = 1) - E(A|T = 0)$. Suppose we found some effect of treatment $T$ on our measure $A$. Once again two possibilities might arise. It might be that $T$ exerts an effect on the underlying latent $\eta$ which affects the indicators $(X_1, \ldots, X_d)$ and thus also our measure $A$ as in Figure 3a. Alternatively, however, it may be the case that $T$ in fact directly affects the indicators $(X_1, \ldots, X_d)$ and thus also our measure $A$ as in Figure 3b. Once again, both of these causal structures are entirely consistent with the measurement model in equation (1) and Figure 1, as indeed are many others, such as if $T$ affected both the latent variable $\eta$ and the indicators $(X_1, \ldots, X_d)$.

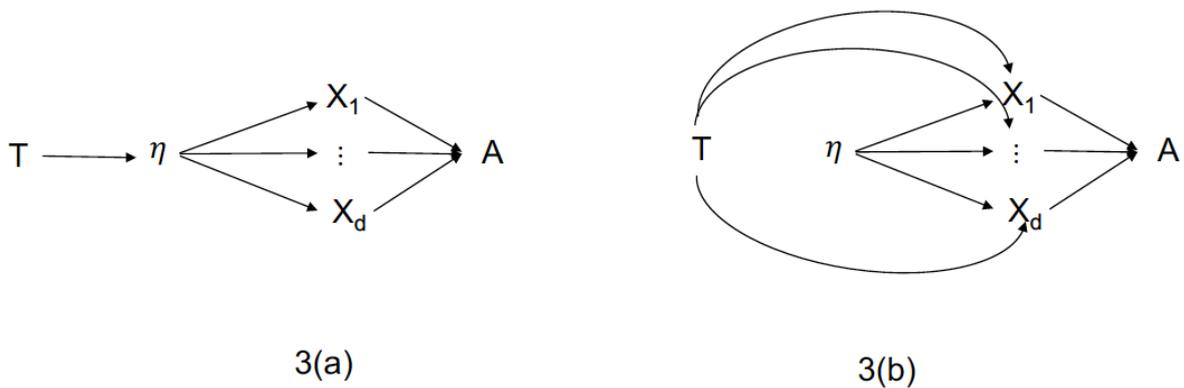

Fig. 3. (a) Structural latent factor model with latent $\eta$ causally affected by treatment $T$; (b) basic latent factor model with indicators $(X_1, \ldots, X_d)$ causally affected by treatment $T$.

In practice, once a measure is formed, with some evidence that the covariance structure amongst the items $(X_1, \ldots, X_d)$ is unidimensional, it is then subsequently presumed that the underlying latent is causally efficacious and thus that the measure $A = f(X_1, \ldots, X_d)$ is reasonably suitable for empirical research intended to examine causal relationships and that the indicators themselves can effectively be ignored once they are used to form the measure $A$. This is what is implicitly assumed when the measure $A$ is used in regression analyses or in forming a typical structural equation model (Bollen, 1989; Sánchez, 2005). But this is presumption. There is nothing in the measurement model in equation (1) fitting well that implies that Figure 2a, rather than 2b or some other model, represent causal relationships with outcome $Y$, or that Figure 3a, rather than 3b or some other, represent causal relationships with treatment $T$.

The assumption that it is the latent variable, rather than the indicators, that is causally efficacious, is precisely that – an assumption. It is an assumption that might be made about the latent variable and the indicators that fit model (1), but it is not one that necessarily holds. To acknowledge that the assumption is not implied by the measurement model in equation (1) and Figure 1, it might be preferable to refer to the models with, and without, the assumption differently. We might refer to the measurement model represented in equation (1) and Figure 1 as the *basic* univariate latent factor model. In contrast, we will refer to the univariate latent factor model as *structural*, if it is further assumed that the indicators, $(X_1, \ldots, X_d)$, do not have causal effects on anything subsequent, and if moreover they are themselves only affected by antecedents through the latent variable $\eta$, so that on a causal diagram (Pearl, 2009), there are no arrows directly out of $(X_1, \ldots, X_d)$, nor into $(X_1, \ldots, X_d)$ except from $\eta$ (and possibly from correlated error terms in models that allow these). On a causal diagram interpreted as a set of non-parametric structural equations with faithfulness (Pearl, 2009) this assumption is equivalent to it being the case that for any other variable $Z$ on the diagram we have that $Z$ is independent of $(X_1, \ldots, X_d)$ conditional on $\eta$. This is a strong assumption. However, it is one that is often effectively implicitly made.

In practice, on the basis of equation (1) fitting the data well, it is often simply assumed that the basic latent factor model is also *structural*, but this is, once again, an assumption. Moreover, it is a strong assumption and one that has empirically testable implications, even though the latent $\eta$ is itself unobserved. Specifically, we show in the Appendix that if the structural latent factor model holds, then the empirical conditions stated in the Theorem below must hold between the indicators $(X_1, \ldots, X_d)$ and any other variable $Z$.

*Theorem 1*. Suppose that $Z$ is independent of $(X_1, \ldots, X_d)$ conditional on $\eta$ and that the basic latent factor model in equation (1) holds, then for any $i$ and $j$, and any values $z$ and $z^*$, we must have $\lambda_i \{ E(X_j | Z = z) - E(X_j | Z = z^*) \} = \lambda_j \{ E(X_i | Z = z) - E(X_i | Z = z^*) \}$.

Theorem 1 has empirical implications provided both $\lambda_i$ and $\lambda_j$ are non-zero. As noted above, when latent factor models are used in practice, it is generally assumed that all $\lambda_i$ are non-zero since any indicator with $\lambda_i = 0$ would typically be discarded. Although in practice it is typically assumed that the causal structure for the latent factor model corresponds to Figure 1, all that is needed to derive the result in Theorem 1 is that equation (1) holds along with the conditional independence of $Z$ and $(X_1, \ldots, X_d)$ given $\eta$. The result is thus applicable even if the indicators $(X_1, \ldots, X_d)$ potentially causally affect one another. The result also holds even if $\varepsilon_i$ in equation (1) is degenerate for one or more indicators so that $X_i = \lambda_i \eta$. As discussed in the Appendix, generalizations are also possible to settings with a multi-dimensional latent variable $\eta$. However, the assumption of a univariate latent variable $\eta$ covers a large number of proposed scales in the psychosocial empirical literature and will be the focus here.

It is easiest to see the implications of this result when the variable $Z$ corresponds to some randomized treatment $T$ and the indicators have been centered. If we apply the result with $Z=z$ corresponding to $T=1$ and $Z=z^*$ corresponding to $T=0$ we obtain:

$$\lambda_i \{ E(X_j | T = 1) - E(X_j | T = 0) \} = \lambda_j \{ E(X_i | T = 1) - E(X_i | T = 0) \}$$

or

$$\{E(X_j|T=1) - E(X_j|T=0)\}/\lambda_j = \{E(X_i|T=1) - E(X_i|T=0)\}/\lambda_i.$$

Essentially what we have is that the effect of the randomized treatment $T$ on $X_j$, scaled by its reliability $\lambda_j$ in model (1) must be the same as the effect of treatment $T$ on $X_i$, scaled by its reliability $\lambda_i$. Intuitively, this must be the case, because under the structural interpretation, $T$ can only act on $X_i$ and $X_j$ through its effects on the latent variable $\eta$. The *structural* latent factor model itself thus has empirical implications in a randomized trial of some treatment $T$. As an extreme example, if in a randomized trial of treatment $T$, it were known that $T$ had an effect on $X_i$ but no effect on $X_j$, then we would know that the structural latent factor model was false, even if the basic latent factor model in (1) would otherwise fit the data well. In other words, if a randomized treatment had an effect on $X_i$ but not on $X_j$ then this would imply that Figure 3b or some other model, rather than 3a, was the correct representation of the causal relations.

While the intuition may be clearest in a randomized trial, in fact similar results apply also if we examine the relationships between the indicators and outcomes, rather than between the indicators and treatments. Suppose we have some binary outcome $Y$ (say, death during follow-up) and that instead of considering treatment values *T=1* and *T=0* we consider outcome values *Y=1* and *Y=0*. By the same logic we would have $\{E(X_j|Y=1) - E(X_j|Y=0)\}/\lambda_j = \{E(X_i|Y=1) - \{E(X_i|Y=0)\}/\lambda_i$. If we compare the prior retrospective values of indicator $X_j$ among those who die during follow-up (*Y=1*) versus survive (*Y=0*) and scale this difference by dividing by the reliability $\lambda_j$ then we should obtain the same quantity as we obtain if we do this with a different indicator $X_i$. The intuition here is that if the *structural* latent factor model holds so that it is the latent $\eta$, rather than the indicators, that have effects on the outcome $Y$, then the latent factor model constrains the relationships between $Y$ and each indicator $X_i$. If these constraints do not hold, then the structural latent factor model cannot be correct, and causal relationships represented by Figure 2b, rather than by Figure 2a, may be more plausible, or it may be the case that both the latent $\eta$ and the indicators have causal effects on the outcome.

While the intuition here corresponds to the causal effect of the latent $\eta$, versus the indicators $(X_1, \dots, X_d)$ on the outcome $Y$, the observed associations between $(X_1, \dots, X_d)$ and $Y$ in fact need not be causal with respect to either $\eta$ or $(X_1, \dots, X_d)$ for Theorem 1 to hold, or for the logic to be applicable. For example, even if, as in Figure 4, the effect of $\eta$ on $Y$ was confounded by covariates $C$, and $C$ was not controlled for in the analysis it is still nevertheless the case that the empirical relations in Theorem 1, with $Z=Y$, must still hold if the structural latent factor model is true. One need not adjust for $C$ to render the empirical relations of Theorem 1 to be necessary under the structural interpretation. Essentially, a structural latent factor model implies not only that causal associations will respect the factor model structure, but also that confounded associations will likewise respect the factor model structure. It is, however, also the case that, under the structural latent factor model, a conditional analogue to Theorem 1 (see the Appendix) likewise applies, wherein every expression in Theorem 1 is conditional on $C$, so that one could alternatively examine equalities conditional on $C$, such as $\{E(X_j|Y=1,c) - E(X_j|Y=0,c)\}/\lambda_j = \{E(X_i|Y=1,c) - \{E(X_i|Y=0,c)\}/\lambda_i$ if one were specifically interested in assessing constraints concerning causal effects. However, again, controlling for a sufficient set of confounding variables for the effect of $\eta$ on $Y$ is not necessary for the empirical equalities to be required to hold under the structural factor model.

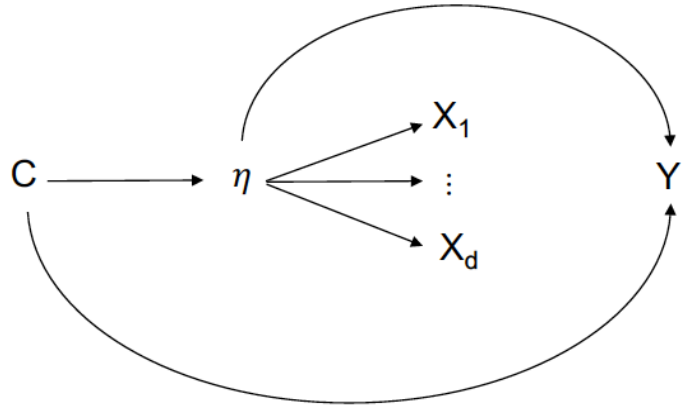

Fig. 4. Causal effect of latent $\eta$ on outcome $Y$ confounded by covariates $C$, but the structural interpretation still requiring indicators $(X_1, \ldots, X_d)$ independent of $Y$ conditional on $\eta$.

Likewise, while the intuitions given above relate to the "effects of randomized treatment on the latent $\eta$" or the "effect of the latent $\eta$ on some other outcome", in fact, under a structural latent factor model, the empirical relations in Theorem 1 must hold for *any other* variable $Z$. Thus, in the causal diagram given in Figure 5, under a structural latent factor model for $\eta$, the empirical relations given in Theorem 1 would have to hold between $(X_1, \ldots, X_d)$ and $Z$, even though $Z$ has no effect on $\eta$, and $\eta$ has no effect on $Z$. Again, under a structural latent factor model, the empirical relations in Theorem 1 must hold between $(X_1, \ldots, X_d)$ and any other variable $Z$.

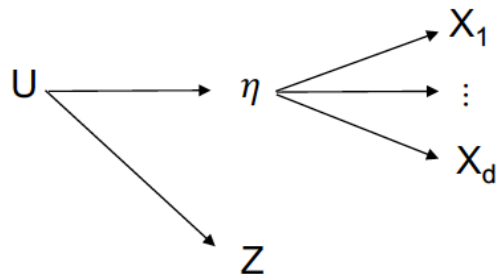

Fig. 5. Structural latent factor model with $Z$ neither affecting, nor affected, by latent $\eta$, but the structural interpretation still requiring indicators $(X_1, \ldots, X_d)$ independent of $Z$ conditional on $\eta$.

We can see then that the assumption that the latent factor model is structural is a very strong assumption. That only the supposedly underlying latent variable $\eta$, and not the indicators, are causally efficacious has numerous testable empirical implications, even though the latent variable $\eta$ itself is unobserved. As noted above, it should be remembered that if the empirical constraints implied by a univariate structural latent factor model do not hold, this does not necessarily imply that the causal models in Figures 2(b) or 3(b) are applicable. The model may be more complex still, potentially involving effects from, or to, the latent variable and the indicators, and possibly involving multiple potentially related latent variables, or none at all (Silva et al., 2006; Kummerfeld et al., 2014; Gignac, 2016; Mansolf and Reise, 2017). Rejection of the univariate structural latent factor model does not settle the question of the appropriate underlying causal structure and further methodology on model selection and identification may be useful in these settings (Bollen, 1989; Spirtes et al., 2000; Silva et al., 2006; Kummerfeld and Ramsey, 2016).

## 3. Statistical Tests to Reject the Structural Latent Factor Model

In developing formal statistical tests of the empirical implications of a structural interpretation of the latent factor model, as expressed by Theorem 1, we will consider two distinct tests. The first test relies on estimates of the reliabilities $\lambda_i$ of the different indicators, which we discuss in Section 3.1 and the test will then be proposed in Section 3.2. This test requires data on at least 3 indicators (in order to identify the reliabilities), but imposes no restrictions on the variable Z in Theorem 1. Reliance on estimates of the reliabilities, however, renders this test dependent on the specific distributional assumptions that are made in fitting model (1), such as assumptions about the mutual independence of the error terms. We view this as potentially undesirable when assessing the supposed structural interpretation of the latent factor model. Thus, in Section 3.3, we introduce an alternative test which does not rely on estimates of the reliabilities. It evaluates exclusively the structural interpretation of the latent factor model, as expressed in terms of its implications given in Theorem 1, making no use of the additional distributional assumptions contained in model (1).

### 3.1 Estimation of reliability

In order to attain a relatively simple testing procedure in Section 3.2, we will make use of reliability estimates that have a simple asymptotic distribution and are easy to obtain with standard software, while being reasonably efficient. We will assume model (1) holds, along with the default assumption that $\eta$ is a mean zero variable with unit variance, independent of the residual errors $\varepsilon_i$, and that the residual errors $\varepsilon_i$ are independent of one another. This then implies that

$$Cov(X_i, X_j) = \lambda_i \lambda_j,$$

for $i,j=1,...,d$. Let $X_{ik}$ denote the value of the indicator $X_i$ across subjects $k=1,...,N$. Consistent estimators $\hat{\lambda}_i$ of $\lambda_i$ for $i=1,...,d$ are thus readily obtained by solving $d$ unbiased estimating equations

$$\sum_{j=1, j\neq i}^{d} \frac{\lambda_j}{N} \sum_{k=1}^{N} (X_{ik} - \bar{X}_i)(X_{jk} - \bar{X}_j) - \lambda_i \lambda_j = \sum_{j=1, j\neq i}^{d} \lambda_j (C_{ij} - \lambda_i \lambda_j) = 0,$$

for $\lambda_i, i = 1, ..., d$, where $\bar{X}_i$ is the sample average of indicator $X_{ik}$ across subjects $k=1,...,N$, and $C_{ij}, i, j = 1, ..., d, i \neq j$, denotes the sample covariance between $X_{ik}$ and $X_{jk}$. When the reliabilities are positive – which is generally plausible, possibly pending reversal of the coding of some indicators so as to render the sample covariances between all pairs of indicators positive - these solutions are readily obtained using off-the-shelf software as the exponentiated estimated coefficients from a quasi-Poisson model with log link, with outcome given by the *d(d-1)*-dimensional vector of sample covariances, without intercept, and with $d$ covariates given by the dichotomous indicators that the considered covariance involves item $X_i, i=1,...,d$. In particular, the solutions $\hat{\lambda}_s$ for $\lambda_s, s = 1, ..., d,$ to the above equations can be calculated as $exp(\widehat{\Lambda}_s)$, with $\widehat{\Lambda}_s$ the maximum pseudo-likelihood estimator of $\Lambda_s$ under working model

$$\log E(C_{ij}) = \sum_{s=1}^{d} \Lambda_s I(s \in \{i, j\}),$$

assuming mutually independent and homoscedastic errors, where $I(.)$ denotes an indicator function, which is 1 if the argument is true and 0 otherwise.

## 3.2 A statistical test dependent on reliability estimates

Consider a discrete $Z$ with categories $\{1,...,p\}$. If the reliabilities $\lambda_i$ of the different indicators were known and non-zero, then a statistical test of the identities in Theorem 1 would be equivalent to a test of the null hypothesis that $\{E(X_j|Z=z) - E(X_j|Z=1)\}/\lambda_j = \{E(X_i|Z=z) - E(X_i|Z=1)\}/\lambda_i$ for $z=2,...,p$, and $i,j=1,...,d$. To test this hypothesis, we will first infer estimators of the conditional expectations $E(X_i|Z=z)$, which obey this null hypothesis. If we define $\gamma_i = E(X_i|Z=1)$ and $\beta_w = \{E(X_1|Z=w) - E(X_1|Z=1)\}$ then under the null hypothesis we can parameterize $E(X_i|Z=z)$ as:

$$E(X_i|Z=z) = \gamma_i + \frac{\lambda_i}{\lambda_1} \sum_{w=2}^{p} \beta_w I(z=w) \tag{2}$$

for $i=1,...,d$, where $\gamma_i, i=1,...,d$ and $\beta_w, w=2,...,p$ are unknown. Let $U_k$ be a $(p \times d)$-dimensional vector with elements $I(Z_k = z)\left\{X_{ik} - \gamma_i - \frac{\lambda_i}{\lambda_1}\sum_{w=2}^{p}\beta_w I(z=w)\right\}$ for $z=1,...,p$, and $i=1,...,d$. With known fixed $\lambda_i, i=1,...,d$, consistent generalized methods of moments estimators (Newey and McFadden, 1994) for $\gamma_i, i=1,...,d$, and $\beta_w, w=2,...,p$ under model (2) are readily obtained by minimizing

$$T_0 = N\left(\frac{1}{N}\sum_{k=1}^{N} U_k^T\right)\Sigma^{-1}\left(\frac{1}{N}\sum_{k=1}^{N} U_k\right),$$

with respect to these parameters, where $\Sigma$ is the empirical covariance matrix of $U_k, k=1,...,N$, and $U_k^T$ is the transpose of $U_k$. It follows from general results on distance metric statistics in Theorem 9.2 of Newey and McFadden (1994) that the resulting minimum serves as a test statistic of the above null hypothesis, which converges to a $\chi^2$ distribution with $(d-1)(p-1)$ degrees of freedom under that hypothesis as the sample size $N$ goes to infinity, provided that the vectors of indicators $(X_{1k},...,X_{dk})$ measured for different subjects are independent and identically distributed and that $\lambda_1$ is non-zero. We refer to the Appendix for justification of the degrees of freedom. Optionally, numerical minimization of $T_0$ can be initiated at inefficient consistent estimators for $\gamma_i$ and $\beta_w$, which can be obtained by fitting model (2) at the given $\lambda_i, i=1,...,d$, using independence generalized estimating equations.

In practice, $\lambda_i, i=1,...,d$, are unknown and must be substituted by estimates $\hat{\lambda}_i$, as given in Section 3.1. To accommodate the uncertainty in these estimated reliabilities in the above test statistic, it suffices to redefine $\Sigma$ in the expression for $T_0$ as the empirical covariance matrix of

$$U_k - \left(\frac{1}{N}\sum_{k=1}^{N}\frac{\partial U_k}{\partial \lambda}\right)\left(\frac{1}{N}\sum_{k=1}^{N}\frac{\partial V_k}{\partial \lambda}\right)^{-1} V_k$$

(see the Appendix), where $\lambda$ is the vector of reliabilities and $V_k$ is a $d$-dimensional vector with elements

$$V_{ik} = \sum_{j=1, j\neq i}^{d} \lambda_j\{(X_i - \bar{X}_i)(X_j - \bar{X}_j) - \lambda_i\lambda_j\},$$

for $i=1,...,d$, as defined in Section 3.1; in this expression, $\lambda_i$ can be substituted by $\hat{\lambda}_i$. As before, consistent generalized methods of moments estimators for $\gamma_i, i=1,...,d$, and $\beta_w, w=2,...,p$ under model (2) are then obtained by minimizing $T_0$, now using the revised choice of $\Sigma$, and the minimized value of $T_0$ null hypothesis can be used as a test statistic of the null hypothesis that the structural interpretation of the latent factor model holds. It converges to a $\chi^2$ distribution with *(d-1)(p-1)* degrees of freedom under that null hypothesis as the sample size *N* goes to infinity, provided that the vectors of indicators *($X_{1k}$,…,$X_{dk}$)* measured for different subjects are independent and identically distributed and that $\lambda_1$ is non-zero.

### 3.3 A statistical test independent of reliability estimates

That the proposed statistical test relies on reliability estimates may be viewed as not entirely satisfactory, because the test may be biased when the covariance structure between the residual error terms in model (1) is misspecified. This could in turn lead to an inflated risk of falsely rejecting the structural interpretation of the latent factor model due solely to the misspecified covariance. In this Section, we will therefore propose an alternative test, which does not rely on reliability estimates and can therefore be used even when only 2 indicators are available, but does then require that *Z* have at least 3 levels. The proposed test is based on the following, equivalent formulation of the identity in Theorem 1.

*Theorem 2*. Under the basic latent factor model in equation (1), for any discrete *Z*, the following two conditions are equivalent:
1) for any *i* and *j*, and any values *z* and $z^*$, $\lambda_i\{E(X_j|Z=z) - E(X_j|Z=z^*)\} = \lambda_j\{E(X_i|Z=z) - E(X_i|Z=z^*)\}$;
2) for any *i* and any values *z*, and for an arbitrary fixed value $z^*$, $E(X_i|Z=z) - E(X_i|Z=z^*) = \alpha_i \beta_z$ for some parameters $\alpha_i, \beta_z$.

Provided at least two of the indicators have non-zero $\lambda_i$ and *Z* has at least 3 levels, we can use this result to construct a test of the empirical implications of the structural latent factor model. Consider again a discrete *Z* with categories *{1,…,p}*. Theorem 2 then suggests that under the null hypothesis that the structural latent factor model holds, we can parameterize $E(X_i|Z=z)$ as:

$$E(X_i|Z=z) = \gamma_i + \sum_{w=2}^{p} \alpha_i \beta_w I(z=w)$$

(3)

for all *i* and all values *z*. This model has *(2d+p-2)* unknown parameters (since the product $\alpha_i \beta_z$ is invariant to rescaling of the form $\alpha_i/\tau$ and $\beta_z\tau$ for any non-zero $\tau$).

A test statistic of the null hypothesis can next be constructed by redefining $U_k$ to be the *(d×p)*-dimensional column vector with elements $I(Z_k=w)(X_{ik} - \gamma_i - \alpha_i\beta_w)$ for *i=1,…,d* and *w=1,…,p*, where we define $\beta_1 = 0$. Consistent generalized methods of moments estimators (Newey and McFadden, 1994) for $\gamma_i, \alpha_i, i = 1, …, d$, and $\beta_w, w = 2, …, p$ under model (3) are readily obtained by minimizing

$$T_1 = N\left(\frac{1}{N}\sum_{k=1}^{N} U_k^T\right) \Sigma^{-1} \left(\frac{1}{N}\sum_{k=1}^{N} U_k\right),$$

where $\Sigma$ is the empirical covariance matrix of $U_k, k=1,…,N$. It then follows from general results on distance metric statistics in Theorem 9.2 of Newey and McFadden (1994) that the resulting minimum is a test statistic of the null hypothesis that the structural latent factor model holds, which converges to a $\chi^2$ distribution with *(d-1)(p-2)* degrees of freedom under the null hypothesis as the sample size $N$ goes to infinity, provided that the vectors of indicators $(X_{1k},...,X_{dk})$ measured for different subjects are independent and identically distributed and that $E(X_i|Z=z)$ is non-zero for at least one $i$ and $z$; we refer to the Appendix for justification of the degrees of freedom and for inefficient consistent estimators that can optionally be used as initial values in this minimization.

The developments in Sections 3.2 and 3.3 consider tests for the null hypothesis that the structural latent factor model holds without making use of covariates $C$. As noted in Section 2, conditioning on $C$ can be done in employing Theorem 1, but such conditioning is optional since the empirical restrictions imposed by the structural latent factor model hold irrespective of such conditioning. With a set of discrete covariates $C$, one could, however, consider conducting tests within strata defined by $C$. In some settings, this may increase power as a result of testing more conditions. However, in other settings this may decrease power due to the reduced sample size in each stratum and the need for multiple testing corrections. Future work will consider extensions to high-dimensional and continuous $C$ and $Z$.

Tests to reject a structural interpretation of a latent factor model could in principle also be carried out using goodness-of-fit tests for structural equation models (Bollen, 1989) comparing, for example, models with arrows from $\eta$ to $Z$, versus models with arrows from $(X_1, ..., X_d)$ to $Z$. This, has not, however, typically been employed in practice. Moreover, we believe the statistical test that is developed here is advantageous over the potential structural equation model approach because (i) the goodness-of-fit test for the structural equation model may also depend on other features of the structural equation model that are not directly relevant to whether it is $\eta$ or $(X_1, ..., X_d)$ that has effects on $Z$; (ii) the statistical test here is applicable under weaker distributional assumptions; and (iii) the test here is applicable for testing the structural interpretation even if both $\eta$ and $(X_1, ..., X_d)$ affect $Z$, or even if neither $\eta$ nor $(X_1, ..., X_d)$ affect, or are affected by, $Z$, as in Figure 5. The test developed here is thus more versatile and is applicable under weaker assumptions. However, as noted above, the test we have developed is targeted to evidence for rejecting the structural interpretation of a univariate latent factor model. Other model selection and identification approaches are important in trying to discern a model that may in fact approximately correspond to the underlying causal structures (Bollen, 1989; Spirtes et al., 2000; Silva et al., 2006; Sullivant et al., 2010; Kummerfeld and Ramsey, 2016).

## 4. Example: The Potential Effect of Life Satisfaction on All-Cause Mortality

Kim et al. (2020) consider the effect of life satisfaction ($\eta$), as assessed by Diener et al.'s (1985) Satisfaction with Life Scale ($A = f(X_1, ..., X_d)$), on subsequent all-cause mortality 4 years later (Y). The Satisfaction with Life Scale (SWLS, Diener et al., 1985) has $d=5$ items, $(X_1, ..., X_d)$, each rated 1-7. These items are: "In most ways my life is close to my ideal"; "The conditions of my life are excellent"; "I am satisfied with my life"; "So far I have gotten the important things I want in life"; and "If I could live my life over, I would change almost nothing." The specific items were chosen and the scale was developed using factor analytic methods. The scale has been documented to have very good psychometric properties: Cronbach's alpha is high and a single underlying factor seems to explain a considerable proportion of the variance across item responses (Diener et al., 1985; Pavot and Diener, 1993). According to Google Scholar the paper that presents the scale (Diener et al., 1985) has now been cited over 34,000 times. In light of the psychometric evidence, the responses to the individual items are thus typically summed for an overall measure, $A = f(X_1, ..., X_d) = \sum_{i=1}^{5} X_i$, between 5 and 35.

The primary analyses of Kim et al. (2020) compared tertiles of this satisfaction of life score in 2010 or 2012 and examined associations with all-cause mortality four years later using data on N=12,998 participants in the Health and Retirement Study, controlling for numerous potentially confounding variables. These included sociodemographic characteristics (age, sex, race/ethnicity, marital status, annual household income, total wealth, level of education, employment status, health insurance, geographic region), childhood abuse, religious service attendance, health conditions and behaviors (diabetes, hypertension, stroke, cancer, heart disease, lung disease, arthritis, overweight/obesity, chronic pain, binge drinking, current smoking status, physical activity, sleep problems), various other aspects of psychological well-being (positive affect, optimism, purpose in life, mastery, depressive symptoms, hopelessness, negative affect, loneliness, social integration), and personality factors (openness, conscientiousness, extraversion, agreeableness, neuroticism). In the primary analysis, those in the top tertile of life-satisfaction were 0.74 (95% CI: 0.64, 0.87) times less likely to die during the four years of follow-up than those in the bottom tertile.

In supplementary analyses, Kim et al. also examined similar associations using each item of Diener et al.'s (1985) Satisfaction with Life Scale separately. Relatively similar risk ratios pertained to 4 of the items: "In most ways my life is close to my ideal" (RR = 0.75; 95% CI: 0.61, 0.91); "The conditions of my life are excellent" (RR = 0.79; 95% CI: 0.66, 0.95); "I am satisfied with my life" (RR = 0.72; 95% CI: 0.62, 0.84); and "So far I have gotten the important things I want in life" (RR = 0.85; 95% CI: 0.73, 0.99). However, for the fifth item "If I could live my life over, I would change almost nothing" the association with all-cause mortality in the four years following was effectively null (RR = 0.98; 95% CI: 0.83, 1.16). There thus appears that there may be some evidence that the indicators of the Satisfaction with Life Scale are differentially associated with all-cause mortality. Here we formally examine whether these data are in fact sufficient to reject the structural interpretation of the latent factor model for the Satisfaction with Life Scale. The Health and Retirement Study data are publicly available online at http://hrsonline.isr.umich.edu/index.php?p=avail&_ga=2.50444521.1751399216.1593436952-1257117760.1593436952; and code for the analysis is available in the Online Supplement.

To investigate the plausibility of the structural interpretation of a latent factor model underlying the scale, we first used the test statistic in Section 3.2 with $Z$ taken to be all-cause mortality at the end of the four-year follow-up. This test statistic requires estimating the reliabilities $\lambda_i$ and the test assumes independence of the residual errors $\varepsilon_i$, an assumption which we will later relax. With $d=5$ indicators and the variable $Z$ taking $p=2$ levels, under the null hypothesis that the factor model is structural, the test statistic $T_0$ in section 3.2 follows a $\chi^2$ distribution with $(d-1)(p-1) = 4$ degrees of freedom. The analysis here again used the Health and Retirement Study data with responses from 12,135 individuals, after removal of incomplete records. After estimating the reliabilities $\lambda_i$ and fitting the restricted model (2), the test statistic $T_0$ was 57.25 after adjusting for the uncertainty in the estimated reliabilities. Comparing this with a $\chi^2$ distribution with 4 degrees of freedom suggests very strong evidence (P = 1.1 ×10$^{-11}$) against the null hypothesis of a structural interpretation of the latent factor model.

To add more robust support to this result, we next evaluated the test statistic of Section 3.3, which does not rely on the distributional assumptions behind model (1), but only on the implications of the structural interpretation of the model. As per Section 3.3, to avoid estimating reliabilities requires a variable $Z$ with at least $p=3$ levels. For this, we let $Z$ be a 4-level variable defined by all-cause mortality and pre-baseline physical activity (measured in 2006 or 2008), coded as 1 (alive and <1x/week of prior vigorous or moderate exercise), 2 (dead and <1x/week of prior vigorous or moderate exercise), 3 (alive and ≥1x/week of prior vigorous or moderate exercise) and 4 (dead and ≥1x/week of prior vigorous or moderate exercise). With

*d=5* indicators and the variable Z taking *p=4* levels, under the null hypothesis that the factor model is structural, the test statistic $T_1$ in section 3.3 follows a $\chi^2$ distribution with *(d-1)(p-2)* = (5-1)(4-2) = 8 degrees of freedom. The analysis here once again used the Health and Retirement Study data with responses from 10,362 individuals, after removal of incomplete records. This resulted in a test statistic $T_1$ of 141.73. Comparing this with a $\chi^2$ distribution with 8 degrees of freedom suggests very strong evidence ($P < 1 \times 10^{-10}$) against the null hypothesis of a structural interpretation of the latent factor model.

That we were able to reject the structural interpretation of the latent factor model for the Satisfaction with Life Scale does *not* imply that it is a bad scale. It arguably does capture well a number of important aspects of a person's satisfaction with the life that he or she has lived, and this is arguably an important outcome to study; the scale is indeed useful for that purpose. Whether absence of regret ("If I could live my life over, I would change almost nothing") ought to be included in that outcome is arguably a conceptual question, concerning the intended coverage of the construct, not an empirical question.

Nevertheless, the rejection here of the structural interpretation of the scale does imply that there is no underlying *univariate* latent variable "life satisfaction" measured by the Satisfaction with Life Scale, such that it is the underlying latent, rather than what is constituted by its indicators, that is causally efficacious. Indeed, different aspects of satisfaction with life appear to be associated with subsequent all-cause mortality in different ways. It may be important to better understand these distinctions and nuances.

One could potentially attempt the formation of a measure that may more closely correspond to an underlying univariate latent with a structural interpretation by dropping the item, "If I could live my life over, I would change almost nothing." This might then render the remaining four items more similarly associated with all-cause mortality. However, this would not necessarily guarantee that comparable associations would still hold with other outcomes (or with the effects on the indicators of various treatments). That would of course require further empirical investigation. However, as noted above and discussed further below, the rejection of the structural interpretation of the latent factor model for the Satisfaction with Life Scale does not mean that the scale ought to be abandoned. It does mean, though, that prior psychometric evidence does not justify such a structural interpretation. Life satisfaction, as assessed by the scale, is not a unidimensional construct with some underlying factor with uniform effects on outcomes.

## 5. Implications and Conclusions

The possibility that the *structural* interpretation of the latent factor model may be wrong – even when the *basic* univariate latent factor model seems to fit the indicators well – has a number of potentially far-reaching implications.

First, the evidence for the structural latent factor model should be established through empirical testing; it should not be presumed. Common practice seems to be to use factor analysis to examine evidence for the uni-dimensionality of a scale. If certain standards and criteria are met, and this is considered established, the scale or measure is then considered "validated" for use in empirical research (DeVellis, 2016; Price, 2016). The scale is employed in other, ideally longitudinal or randomized studies, to examine evidence for causes and effects. However, as this paper has made clear, even if the *basic* univariate latent factor model holds, this does not imply that the interpretation of that model is necessarily *structural* i.e. that it is the supposedly underlying latent factor, rather than the indicators (or whatever phenomena are related to them), that is causally efficacious. It is a big leap to assume that this is so. It is a leap that is often made with no evidence, but such practices could change.

As shown in this paper it is possible to empirically test for, and reject, the implications of a structural interpretation of the univariate latent factor model. It is conceivable that over many outcomes, or in examining the effects of numerous treatments or interventions, none of these tests reject the empirical implications of the structural latent factor model given in Theorem 1. It is possible that the implications of structural latent factor model described in Theorem 1 do closely hold with all outcomes examined, and with all treatments examined. This is not a proof that the structural interpretation holds, but it would constitute evidence. The implications of the structural interpretation of the univariate latent factor model are not fully empirically verifiable, but if numerous tests across numerous different outcomes and treatments did not reject, one might have reason to believe that the structural latent factor model held to a reasonable approximation. One might also use other model selection and identification approaches to compare the univariate structural latent factor model to other more complex models (Bollen, 1989; Spirtes et al., 2000; Silva et al., 2006; Kummerfeld et al., 2014; Kummerfeld and Ramsey, 2016). After carrying out this work, one might eventually be more justified in assuming that the construct under consideration was reasonably well represented by a *univariate* causally efficacious latent variable. One might thus also be more justified in subsequently using the corresponding univariate scale, with other treatments and outcomes in precisely the way that these scales are used at present (but arguably, currently, without adequate justification).

It is entirely possible that in some cases the structural univariate latent factor model will be a reasonably good approximation, while in other cases it will not. But until we examine the evidence, we do not know. Simply showing conformity to the *basic* univariate latent factor model is not sufficient. Again, this tells us nothing about the potential causal efficacy, or not, of the supposed univariate latent variable versus the indicators. Conformity to the basic univariate latent factor model does not help us distinguish between the causal structures in Figure 2a versus 2b or in Figure 3a versus 3b or between other more complex structures. In this regard, not only *exploratory* factor analysis, but also in fact *confirmatory* factor analysis are effectively hypothesis-generating with respect to whether we have identified a univariate factor with a structural causal interpretation. Evidence from exploratory and confirmatory factor analysis that leads to the identification of factors fitting the *basic* latent factor model in equation (1) are a good place to begin with respect to assessing whether that supposed univariate factor is in fact *structural*. But this should be considered the beginning of that process, rather than its conclusion. If the search for factors is ultimately to uncover constructs that can be effectively represented by a univariate and causally efficacious latent variable in the structural sense, then it will subsequently be desirable to evaluate whether the empirical relations given in Theorem 1 hold with respect to a range of treatments and outcomes. Evidence for so-called measurement invariance (Meredith, 1993; Cheung and Rensvold, 2002; Meredith and Teresi, 2006; Putnick and Bornstein, 2016) might help partially mitigate the possibility of differential treatment effect across the indicators, but does nothing to ensure comparable associations of the indicators with other subsequent outcomes. In any case, the implications of the *structural* univariate factor model should be examined to justify current practices of using univariate scales in ensuing subsequent empirical research.

A second related implication also follows from this: until there is substantial evidence already in place, from multiple outcomes and multiple treatments, that the *structural* factor model holds, we should, until that evidence is established, continue to examine the potential casual relationships between individual indicators and outcomes, and between treatments and individual indicators, one indicator at a time. The current practice is that once it is shown that the *basic* latent factor model fits a set of indicators well, and the items and scale meet other criteria (DeVellis, 2016), then, it is typically assumed that this is adequate justification for using the scale, and not the individual indicators, in all subsequent research. Indeed, investigators are not infrequently criticized for the practice of examining associations between outcomes and individual indicators. However, if the structural interpretation of the univariate latent factor model has not been established, then such criticisms are inappropriate. Indeed item-by-item examination may be important for uncovering the nuances of the constructs being considered and their differing relations to outcomes of interest (VanderWeele, 2022).

It is entirely possible for the basic factor model to hold with a multi-item scale and yet, for example, to have only one of the indicators have any causal efficacy. Such would be the case in Figure 6a with only the item $X_1$ having causal efficacy or in Figure 6b in which each indicator $X_i$ corresponds to some underlying latent $\eta_i$ but only $\eta_1$ has causal efficacy for the outcome $Y$.

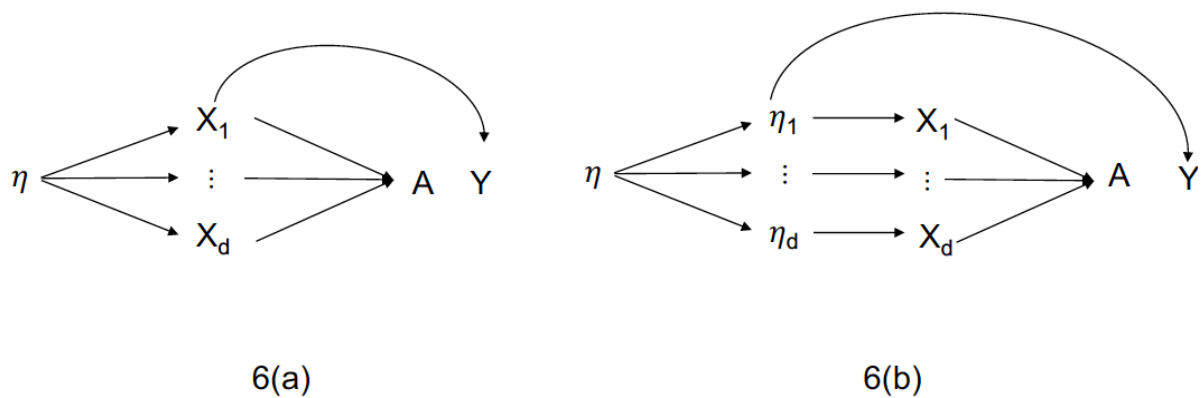

6(a)  6(b)

Fig. 6. Basic latent factor model for $\eta$ with only (a) one single indicator $X_1$, or (b) one single subsequent latent $\eta_1$, causally efficacious for outcome $Y$.

Both of these models in Figure 6a and 6b are entirely consistent with the basic factor model perfectly fitting the data for the set of indicators $(X_1, \ldots, X_d)$. If we do not examine the indicators' relationships one at a time with the outcome we would miss this critical nuance. If the indicators are strongly correlated with one another, then we will still see substantial association between the measure $A$ (constructed by e.g. an average of the indicators) and the outcome $Y$, but we will not see that this is attributable solely to e.g. $X_1$ in Figure 6a (or that which underlies it, $\eta_1$, in Figure 6b). To see this, we would need to regress $Y$ on all of the indicators $(X_1, \ldots, X_d)$ simultaneously, and under the structure in Figure 6a, for example, we would then see that only $X_1$ is relevant for the outcome.

In many cases, we may find that the structural interpretation of univariate latent factor model does not hold, as was the case for the Satisfaction with Life Scale. This does not mean the scale should be abandoned. It may be an appropriate or desired summary of a set of indicators or items, interpreted simply as an average of these. It may be useful as an outcome. The scale may also potentially be used as an exposure or independent variable of interest, even if the structural interpretation does not hold, as it is still possible to give the estimates using the scale a causal interpretation, albeit one that is more nuanced using causal inference theory for multiple versions of treatment (VanderWeele and Hernán, 2013; VanderWeele, 2022). However, while we can still use a scale, even as a single univariate exposure without having established the structural interpretation of the underlying factor model, we should keep in mind that we may be obscuring important distinctions and differential relationships across indicators. If we do not already have considerable evidence for the structural interpretation then we certainly should not criticize item-by-item analyses. Indeed, these may be precisely what is helpful in gaining more nuanced insight.

Third, if the latent factor model is not in fact structural, then an intervention itself can alter the observed factor structure across items. Suppose that for a set of indicators $(X_1, ..., X_d)$ a univariate latent factor model as in equation (1) fits the data well. Suppose that a new treatment $T$ is introduced that affects indicators $X_1$ and $X_2$ but none of the other indicators $(X_3, ..., X_d)$ as in Figure 7.

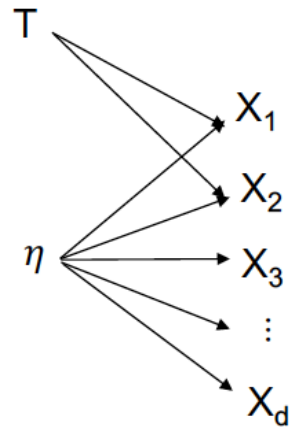

Fig. 7. Basic latent factor model for $\eta$ with treatment $T$ directly affecting indicators $X_1$ and $X_2$, and thereby altering the factor structure.

Suppose further that over time the use of treatment $T$ becomes more widespread so that it would be present among many individuals in most samples. While a univariate latent factor model might have originally fit the data well, once $T$ is introduced, the factor structure has been changed. A new treatment may alter the factor structure. This cannot happen if the latent factor model is structural so that any and all effects on $(X_1, ..., X_d)$ operate through $\eta$, but if the latent factor model is basic but not structural, then treatments can change the properties of the basic factor model.

Fourth, taking these various possibilities into account, there is a potentially dangerous process of feedback that has occurred between (i) numerous statistical analyses often suggesting that, in many settings, a basic univariate latent factor explains reasonably well the covariance across item responses; (ii) the supposition that this then entails a structural interpretation of the latent factor model; (iii) the fact that causal relationships between latent factors can suggest a basic univariate latent factor model, even when the true underlying structures are multivariate (VanderWeele and Batty, 2020); and (iv) the lack of consideration of causal relations between underlying structural latent factors, or between prior causes and indicators leading to an overreliance on factor analysis with one wave of data (VanderWeele and Batty, 2020). That the covariance structures across indicators does, in scale development, often seem to fit well a one-factor model reinforces the supposition that this is nothing unusual and then further reinforces the fundamental error of concluding that the univariate latent factor model is structural. In light of the considerations above, current practices of factor analysis do not constitute an adequate approach for model selection or for scale development.

These mutually reinforcing experiences and suppositions need to be revisited. Differential associations of the indicators in well-established scales with specific outcomes (or specific treatments, or other variables) and the tests in this paper will challenge the *structural* interpretation of latent factor models. The recognition that these latent factor models may not be structural, even if the *basic* univariate latent factor model fits the data well, might lead to a better appreciation that there may be important aspects of the constructs under consideration that are multidimensional in nature, and that these various dimensions may be very differently related to outcomes of interest. It may then also become apparent that there has been an overreliance on factor analysis itself for model selection and for scale development and other approaches may be considered (Spirtes et al., 2000; Silva et al., 2006; Sullivant et al., 2010; Kummerfeld and Ramsey, 2016). It is time that this process of re-examination begins. We may need to return to well-established scales and consider the items one-by-one, and utilize the tests described in this paper and employ further tests and methods for more complex settings, to evaluate the evidence pertaining to whether well-fitting univariate latent factor models are indeed structural, or whether the presumption that they are, has obscured important distinctions.

**Appendix**

Here we discuss generalizations of Theorem 1 that allow for conditioning on a set of covariates $C$ and that may involve a multi-dimensional latent variable $\eta$.

*Generalization of Theorem 1.* Suppose that $Z$ is independent of $(X_1, \ldots, X_d)$ conditional on univariate $\eta$ and covariates $C$, and that the basic latent factor model in equation (1) holds such that $X_i = \lambda_i \eta + \varepsilon_i$ with $\varepsilon_i$ independent of $\eta$ conditional on $C$, then for any $i$ and $j$, and any values $z$ and $z^*$, we must have $\lambda_i \{E(X_j|Z = z, c) - E(X_j|Z = z^*, c)\} = \lambda_j \{E(X_i|Z = z, c) - E(X_i|Z = z^*, c)\}$.

*Proof of Generalization of Theorem 1.* For any variable such that $Z$ is independent of $(X_1, \ldots, X_d)$ conditional on $(\eta, C)$ we have that

$$E(X_i|Z = z, c) = E_\eta\{E(X_i|Z = z, \eta, c)|Z = z, c\}$$
$$= E_\eta\{E(X_i|\eta, c)|Z = z, c\}$$
$$= E_\eta\{\lambda_i \eta + E(\varepsilon_i|c)|Z = z, c\}$$
$$= \lambda_i E_\eta(\eta|Z = z, c) + E(\varepsilon_i|c).$$

It likewise follows that

$$E(X_i|Z = z^*, c) = \lambda_i E_\eta(\eta|Z = z^*, c) + E(\varepsilon_i|c),$$

from which

$$E(X_i|Z = z, c) - E(X_i|Z = z^*, c) = \lambda_i \{E_\eta(\eta|Z = z, c) - E_\eta(\eta|Z = z^*, c)\}.$$

Let us first assume that $\lambda_i$ differs from 0. Then $E_\eta(\eta|Z = z, c) - E_\eta(\eta|Z = z^*, c) = \{E(X_i|Z = z, c) - E(X_i|Z = z^*, c)\}/\lambda_i$. Now applying this result again to $E(X_j|Z = z, c)$ we similarly obtain $E_\eta(\eta|Z = z, c) - E_\eta(\eta|Z = z^*, c) = \{E(X_j|Z = z, c) - E(X_j|Z = z^*, c)\}/\lambda_j$, provided that also $\lambda_j$ differs from 0, and thus we must have $\lambda_i\{E(X_j|Z = z, c) - E(X_j|Z = z^*, c)\} = \lambda_j\{E(X_i|Z = z, c) - E(X_i|Z = z^*, c)\}$. When $\lambda_i = 0$, then it follows from the earlier derivation that $E(X_i|Z = z, c) - E(X_i|Z = z^*, c) = 0$, so that the identity continues to hold. When the set of covariates $C$ is empty we obtain the result given in the text. This completes the proof.

Suppose now that $\eta$ is multivariate and that a basic latent factor model holds such that $X_i = \lambda_i' \eta + \varepsilon_i$ with $\varepsilon_i$ independent of $\eta$ conditional on $C$, with $\lambda_i$ constituting a vector. Suppose once again that $Z$ is independent of $(X_1, \ldots, X_d)$ conditional $\eta$ and covariates $C$. Let $X = (X_1, \ldots, X_d)'$ and $\Lambda = (\lambda'_1, \ldots, \lambda'_d)$ and $\epsilon = (\epsilon_1, \ldots, \epsilon_d)'$ so that $X = \Lambda\eta + \epsilon$. We then have that

$$E(X|Z=z,c) = E_\eta\{E(X|Z=z,\eta,c)|Z=z,c\}$$
$$= E_\eta\{E(X|\eta,c)|Z=z,c\}$$
$$= E_\eta(\Lambda\eta|Z=z,c) + E(\epsilon|c)$$
$$= \Lambda E_\eta(\eta|Z=z,c) + E(\epsilon|c)$$

from which it follows

$$E(X|Z=z,c) - E(X|Z=z^*,c) = \Lambda\{E_\eta(\eta|Z=z,c) - E_\eta(\eta|Z=z^*,c)\}$$

This set of $d$ equations will in general give us empirical implications from the observed data provided $d > dim(\eta)$.

*Proof of Theorem 2.* Let us denote $E(X_i|Z=1) = \gamma_i$. For indicators, $i$ and $j$, the basic identity in Theorem 1 is that

$$\lambda_j\{E(X_i|Z=z) - \gamma_i\} = \lambda_i\{E(X_j|Z=z) - \gamma_j\}$$

for all $i,j,z$. Consider the analogous expression evaluated at some $z^*$. Multiplying both expressions, with $j=1$, shows that

$$\lambda_1\lambda_i\{E(X_i|Z=z^*) - \gamma_i\}\{E(X_1|Z=z) - \gamma_1\} = \lambda_i\lambda_1\{E(X_i|Z=z) - \gamma_i\}\{E(X_1|Z=z^*) - \gamma_1\},$$

and, under the assumption that $\lambda_i$ is non-zero for all $i$, that

$$\{E(X_i|Z=z^*) - \gamma_i\}\{E(X_1|Z=z) - \gamma_1\} = \{E(X_i|Z=z) - \gamma_i\}\{E(X_1|Z=z^*) - \gamma_1\},$$

for all $i,j,z,z^*$. Denote

$$\alpha_i \equiv \{E(X_i|Z=z^*) - \gamma_i\}/\{E(X_1|Z=z^*) - \gamma_1\}$$

and

$$\beta_z \equiv E(X_1|Z=z) - \gamma_1,$$

we conclude that if the identity in Theorem 1 holds, then

$$E(X_i|Z=z) = \gamma_i + \alpha_i\beta_z.$$

If $E(X_1|Z=z^*) - \gamma_1 = 0$, then the result continues to hold upon changing the references category $z^*$ or $j=1$ so that $E(X_1|Z=z^*) - \gamma_1$ is no longer zero; if this is not possible, then $Z$ is independent of all indicators and thus the identity holds trivially.

Conversely, given the reference value $j$ for which $\alpha_j$ differs from zero, the latter identity implies that

$$\lambda_j\{E(X_i|Z=z) - E(X_i|Z=z^*)\} = \lambda_j\alpha_j(\beta_z - \beta_{z^*})\alpha_i/\alpha_j = \lambda_i\alpha_j(\beta_z - \beta_{z^*})$$
$$= \lambda_i\{E(X_j|Z=z) - E(X_j|Z=z^*)\}$$

with

$$\lambda_i = \lambda_j\alpha_i/\alpha_j$$

thus confirming the identity in Theorem 1. Note that such reference value *j* for which α_j differs from zero can always be found except when $E(X_i|Z = z) = E(X_i|Z = z^*)$ for all *i* and *z*, in which case the identity in Theorem 1 still holds.

*Asymptotic distribution of the test statistic of Section 3.2.* Define

$$T_0(\lambda) = N \left(\frac{1}{N}\sum_{k=1}^{N} U_k^T(\lambda)\right) \Sigma^{-1} \left(\frac{1}{N}\sum_{k=1}^{N} U_k(\lambda)\right),$$

where we suppress dependence on $\gamma_i, i = 1, \ldots, d$ and $\beta_w, w = 2, \ldots, p$, as these are substituted by the corresponding minimizers of $T_0(\lambda)$ at the given $\lambda$. For fixed, known $\lambda$, it follows by a direct application of Theorem 9.2 in Newey and McFadden (1994) that $T_0(\lambda)$ asymptotically follows a chi-square distribution. For the degrees of freedom, consider the saturated parameterization

$$E(X_i|Z = z) = \gamma_i + \sum_{w=2}^{p} \theta_{iw} I(z = w)$$

The null hypothesis can be expressed as $\lambda_1 \theta_{iz} - \lambda_i \theta_{1z} = 0$ for all *i=1,...,d* and *z=2,...,p*. Let $\theta$ and $a(\theta)$ be vectors of *d(p-1)* elements $\theta_{iz}$ and $\lambda_1 \theta_{iz} - \lambda_i \theta_{1z}$, respectively; we choose not to include $\gamma_i$ in $\theta$ as the null hypothesis does not involve $\gamma_i$, so that the reasoning below would remain unchanged when including it. The vector $a(\theta)$ has *p-1* elements that are guaranteed to equal zero, regardless of the values of $\theta_{iz}$, namely those where *i=1*. Without loss of generality, we place these elements in the first *p-1* rows of $a(\theta)$. The first *p-1* rows in the gradient of $a(\theta)$ with respect to $\theta$ thus contain only zeroes. The remaining rows have $\lambda_1$ on the diagonal position, and 0 on all other positions except possibly the first *p-1* columns. We conclude that the rank of this gradient is *d(p-1)-(p-1)=(d-1)(p-1)* when $\lambda_1$ differs from zero. It then follows from Newey and McFadden (1994) that the proposed test has *(d-1)(p-1)* degrees of freedom.

We next study the asymptotic behaviour of $T_0(\hat{\lambda})$ for the same choice of $\Sigma$. We will later deal with the estimation of $\Sigma$. It follows by a Taylor expansion that

$$T_0(\hat{\lambda}) = T_0(\lambda) + 2 \left(\frac{1}{\sqrt{N}}\sum_{k=1}^{N} U_k^T(\lambda)\right) \Sigma^{-1} \left(\frac{1}{N}\sum_{k=1}^{N} \frac{\partial U_k}{\partial \lambda}(\lambda)\right) \sqrt{N}(\hat{\lambda} - \lambda) + o_p(1).$$

Here, the middle term on the right is uniformly tight but not generally converging to zero in probability because $\frac{1}{\sqrt{N}}\sum_{k=1}^{N} U_k^T(\lambda)$ and $\sqrt{N}(\hat{\lambda} - \lambda)$ are uniformly tight (and not converging to zero in probability), and $\frac{1}{N}\sum_{k=1}^{N} \frac{\partial U_k}{\partial \lambda}(\lambda)$ converges in probability to a non-zero constant. In view of this, we consider

$$T_0^*(\lambda) = N \left(\frac{1}{N}\sum_{k=1}^{N} U_k^{*T}(\lambda)\right) \Sigma^{*-1} \left(\frac{1}{N}\sum_{k=1}^{N} U_k^*(\lambda)\right),$$

where

$$U_k^*(\lambda) = U_k(\lambda) - E\left(\frac{\partial U_k}{\partial \lambda}(\lambda)\right) E\left(\frac{\partial V_k}{\partial \lambda}(\lambda)\right)^{-1} V_k(\lambda),$$

and $\Sigma^*$ is its covariance matrix. For given, known $\lambda$, this follows the same asymptotic distribution as before (by Theorem 9.2 in Newey and McFadden (1994)), provided that $\gamma_i, i = 1, \ldots, d$ and $\beta_w, w = 2, \ldots, p$, are substituted by the corresponding minimizers of $T_0^*(\lambda)$ at the given $\lambda$. By a similar Taylor expansion as before, we then have

$$T_0^*(\hat{\lambda}) = T_0^*(\lambda) + 2 \left(\frac{1}{\sqrt{N}}\sum_{k=1}^{N} U_k^{*T}(\lambda)\right) \Sigma^{*-1} \left(\frac{1}{N}\sum_{k=1}^{N} \frac{\partial U_k^*}{\partial \lambda}(\lambda)\right) \sqrt{N}(\hat{\lambda} - \lambda) + o_p(1),$$

where the middle term on the right now converges to zero in probability since $\frac{1}{N}\sum_{k=1}^{N}\frac{\partial U_k^*}{\partial \lambda}(\lambda)$ converges to zero in probability by the weak law of large numbers and all remaining terms are uniformly tight. We conclude that $T_0^*(\hat{\lambda})$ and $T_0^*(\lambda)$ follow the same asymptotic distribution. The result in Section 3.2 now follows upon noting that, by Slutsky's theorem, the substitution of $\Sigma^*$ by a consistent estimator (which involves a consistent estimator of $\lambda$) does not change the asymptotic distribution.

The above Taylor expansion does not explicate the fact that a change from $\hat{\lambda}$ to $\lambda$ may also change the values for $\gamma_i, i = 1, \ldots, d$ and $\beta_w, w = 2, \ldots, p$, at which the test statistic is minimized. Let $\vartheta$ be the vector of values $\gamma_i, i = 1, \ldots, d$ and $\beta_w, w = 2, \ldots, p$, and $\hat{\vartheta}(\lambda)$ be the corresponding minimizer of $T_0^*(\lambda)$. Then the righthand side of the above Taylor expansion for $T_0^*(\hat{\lambda})$ in principle must include the additional term

$$2\left(\frac{1}{\sqrt{N}}\sum_{k=1}^{N}U_k^{*T}(\lambda)\right)\Sigma^{*-1}\left(\frac{1}{N}\sum_{k=1}^{N}\frac{\partial U_k^*}{\partial \vartheta}(\lambda)\right)\frac{\partial \hat{\vartheta}(\lambda)}{\partial \lambda}\sqrt{N}(\hat{\lambda}-\lambda).$$

We ignored this term since $\left(\frac{1}{\sqrt{N}}\sum_{k=1}^{N}U_k^{*T}(\lambda)\right)\Sigma^{*-1}\left(\frac{1}{N}\sum_{k=1}^{N}\frac{\partial U_k^*}{\partial \vartheta}(\lambda)\right) = 0$ by virtue of $\hat{\vartheta}(\lambda)$ minimizing $T_0^*(\lambda)$, and since all remaining terms are uniformly tight.

*Remark.* Note that $V_k(\lambda)$ does not merely involve the data for subject $k$, but depends on other subjects' data through the sample averages $\bar{X}_i$. This does not pose additional complications because $\{X_i - E(X_i)\}\{X_j - E(X_j)\}$ is the influence function for the covariance between $X_i$ and $X_j$ under the nonparametric model (see e.g. Hines et al. (2022)). Asymptotically equivalent results would therefore be obtained when redefining $V_{ik} = \sum_{j=1, j\neq i}^{d}\lambda_j[\{X_i - E(X_i)\}\{X_j - E(X_j)\} - \lambda_i\lambda_j]$.

*Initial values for the parameters in the minimization of Section 3.3.* In this paragraph, we provide easy-to-calculate, but inefficient consistent estimators of $\gamma_i$ and $\alpha_i$, $i=1,\ldots,d$, and $\beta_w$, $w=2,\ldots,p$, which can optionally be used as starting values in the minimization procedure. In particular, we will estimate $\gamma_i$, $i=1,\ldots,d$, as the solution to

$$0 = \sum_{k=1}^{N} X_{ik} - \gamma_i - \alpha_i\beta_{Z_k},$$

and $\alpha_i$, $i=1,\ldots,d$, as the solution to

$$0 = \sum_{k=1}^{N} \beta_{Z_k}(X_{ik} - \gamma_i - \alpha_i\beta_{Z_k}),$$

where $N$ is the number of subjects. We will moreover estimate $\beta_z$, $z=2,\ldots,p$ by solving

$$0 = \sum_{k=1}^{N} I(Z_k = z)\sum_{i=1}^{d}\alpha_i(X_{ik} - \gamma_i - \alpha_i\beta_z)$$

for $\beta_z$. These unbiased estimating equations can be solved by iterating the following procedure until convergence:

1. Estimate $\beta_z$ as $\hat{\beta}_z = \frac{\sum_{i=1}^{d}\sum_{k=1}^{N}\alpha_i(X_{ik}-\gamma_i)I(Z_k=z)}{\sum_{i=1}^{d}\sum_{k=1}^{N}\alpha_i^2 I(Z_k=z)}$ with $\gamma_i$ and $\alpha_i$ set at the current iteration.

2. Estimate $\alpha_i$ as $\hat{\alpha}_i = \frac{\sum_{k=1}^{N}\beta_{Z_k}(X_{ik}-\gamma_i)}{\sum_{k=1}^{N}\beta_{Z_k}^2}$ with $\gamma_i$ and $\beta_{Z_k}$ set at the current iteration.

3. Estimate $\gamma_i$ as $\hat{\gamma}_i = \frac{\sum_{k=1}^{N}X_{ik}-\alpha_i\beta_{Z_k}}{N}$ with $\alpha_i$ and $\beta_{Z_k}$ set at the current iteration.

As initial estimates of $\gamma_i$ and $\alpha_i$, $i=1,...,d$, in this iterative process, we can use 0 and the sample average of $X_i$ in the subgroup $Z=1$, respectively. We can assess convergence by monitoring the sum of squares of the function values to the above $2d+p-1$ estimating equations with $\gamma_i$, $\alpha_i$ and $\beta_z$ substituted by $\hat{\gamma}_i$, $\hat{\alpha}_i$ and $\hat{\beta}_z$, respectively; this sum of squares should converge to zero.

*Degrees of freedom of the test statistic of Section 3.3.* Consider the saturated parameterization

$$E(X_i|Z=z) = \gamma_i + \sum_{w=2}^{p} \theta_{iw} I(z=w)$$

Reasoning as in the proof of Theorem 2, the null hypothesis can be expressed as $\theta_{12}\theta_{iz} - \theta_{i2}\theta_{1z} = 0$ for all $i=1,...,d$ and $z=2,...,p$. Let $\theta$ and $a(\theta)$ be vectors of $d(p-1)$ elements $\theta_{iz}$ and $\theta_{12}\theta_{iz} - \theta_{i2}\theta_{1z}$, respectively; we choose not to include $\gamma_i$ in $\theta$ as the null hypothesis does not involve $\gamma_i$, so that the reasoning below would remain unchanged when including it. The vector $a(\theta)$ has $d+p-2$ elements that are guaranteed to equal zero, regardless of the values of $\theta_{iz}$, namely those where either $i=1$ or $z=2$. Without loss of generality, we place these elements in the first $d+p-2$ rows of $a(\theta)$. The first $d+p-2$ rows in the gradient of $a(\theta)$ with respect to $\theta$ thus contain only zeroes. The remaining rows have $\theta_{12}$ on the diagonal position, and 0 on all other positions except possibly the first $d+p-2$ columns. We conclude that the rank of this gradient is $d(p-1)-(d+p-2)=(d-1)(p-2)$ when $\theta_{12}$ differs from zero. It then follows from Newey and McFadden (1994) that the proposed test has $(d-1)(p-2)$ degrees of freedom. When $\theta_{12}$ is zero, one may change the coding of the indices $i$ and $z$ to render it non-zero, except when $Z$ is independent of all indicators.

## Acknowledgements

This research was funded by the National Institutes of Health, U.S.A.